\documentclass[prl,twocolumn,showpacs]{revtex4}

\usepackage{amssymb}
\usepackage{bm}

\begin{document}

\title
{Charged vortices in superfluid systems with pairing of spatially
separated carriers}
\author{S.\,I.\,Shevchenko}

\affiliation{%
B.\,I.\, Verkin Institute for Low Temperature Physics and
Engineering National Academy of Sciences of Ukraine, Lenin av. 47
Kharkov 61103, Ukraine}

\date{\today}

\begin{abstract}
 It is shown that in a magnetic field the vortices in superfluid
 electron-hole systems carry a real electrical charge. The
 charge value  depends on the relation between  the
 magnetic length $\ell_B$ and the Bohr radiuses of electrons
 $a_B^e$ and holes $a_B^h$. In double layer systems at filling
 factors $\nu_e=\nu_h=\nu$
 and for $a_B^e,a_B^h\gg \ell_B$
 the vortex charge is equal to the universal value $\nu e$
 .

\end{abstract}


\maketitle

It is generally believed, that the vortices in superconductors are
connected with an applied magnetic field, while the magnetic field
does not have any influence on the properties of the vortices in
electrically neutral superfluid systems. The aim of this letter is
to show that in superfluid systems subjected by a magnetic field
the vortices will have a real electrical charge (the compensating
charge of the opposite sign will appear on the surface of the
system). In general case the charge of the vortices takes a
fractional value. For the first time the fractional charge of the
vortices was predicted by Laughlin \cite{1} for the
two-dimensional electron gas in a quantized magnetic field. Then,
it was established in Ref. \cite{2, 2a}, that in double layer
electron systems with the half-filling of the lowest Landau levels
in each layer the vortices should carry the charge equals to $\pm
e/2$ (here and below $e$ is the absolute value of the electron
charge).

We will show that in any superfluid system the magnetic field
results in an appearance of the vortex charge proportional to the
polarizability of the particles and inverse proportional to their
effective mass. The estimates show that for reachable values of
magnetic fields the vortex charge is unobservable small one in
superfluid phases of He isotopes and in Bose gases of alkali
metals, while it can be of order of the electron charge in
superfluid systems with pairing of spatially separated electrons
and holes.

The authors of Ref. \cite{3a} call the possibility of the
electron-hole pair superfluidity in question based on the fact
that the interband transitions fix the phase of the order
parameter and result in a transition into a dielectric state. But
it was established in Refs. \cite{3,4} that the interdiction on
the electron-hole superfluidity can be removed in systems, where
the spatially separated electrons and holes are coupled. In these
systems the interband transitions coincide with the interlayer
ones and usually they are exponentially small.  The superfluid
state of the  pairs with spatially separated components has both
the superfluid and superconducting features. The superfuid flow of
such electron-hole pairs is accompanied with real supercurrents
flowing in opposite directions. Therefore, we will call  these
systems the condenser superconductors.

In Refs. \cite{3,4} the pairing of a conducting band electron from
the one layer with a valence band hole from the other layer was
considered. Then in a number of theoretical papers
\cite{2,2a,5,6,7,8} it was shown  a possibility of superfluidity
of  pairs composed from spatially separated electrons and holes
belonging to the conducting band. This possibility is realized in
double layer electron systems in a magnetic field normal to the
layers for the case of the total filling factor
$\nu_T=\nu_1+\nu_2=1$. During almost 10 years there were many
efforts to observe the condenser superconductivity experimentally
\cite{9,10,11,12}. Now it seems that these effort have been
crowned with  success \cite{13,14}.

The principal result can be obtained from general consideration
which does not imply a concrete form of the Hamiltonian. Let us
consider a superfluid system subjected with crossed electric ${\bf
E}$ and magnetic ${\bf B}$ fields. Let $\mathcal{ E}$ is the
density of the energy of the system, and $\bm{\Pi} $ is the
density of its generalized momentum. Note, that the values of
$\mathcal{E}$ and $\bm{\Pi}$ do not include the energy and the
momentum of the external fields that polarize the system. We are
interesting in a relation between the momentum density $\bm{\Pi} $
and the dipole momentum density {\bf P}. To establish this
relation it is convenient to consider the frame of reference, in
which the electric field is equal to zero.
The velocity of that
frame relative to the lab frame is equal to
\begin{equation}
{\bf u}= c \frac{{\bf E}\times {\bf B}}{B^2}. \label{1}
\end{equation}
The expression (\ref{1}) is valid to within the linear order of
the factor $u/c$. The condition $u/c\ll 1$ is satisfied at $E^2\ll
B^2$. Below we assume  this condition is fulfilled.

 Let us introduce
the notation $\mathcal{E}_0$ and $\bm{\Pi}_0$, the energy density
and the momentum density in the new frame, correspondingly. The
relation between the energies and the momenta in the lab frame and
in the new one can be obtained from the transformations
\begin{equation}
\bm{\Pi}=\bm{\Pi }_0 + \rho {\bf u} , \label{2}
\end{equation}
\begin{equation}
\mathcal{E}=\mathcal{E}_0(\bm{\Pi }_0)+\bm{\Pi }_0 {\bf u}+
\frac{\rho u^2}{2}. \label{3}
\end{equation}
Here $\rho $ is the mass density. Expressing the momentum $\bm{\Pi
}_0$ in Eq. (\ref{3}) in terms of the momentum $\bm{\Pi }$  we
obtain
\begin{equation}
\mathcal{E}=\mathcal{E}_0(\bm{\Pi }-\rho {\bf u})+ \bm{\Pi } {\bf
u}- \frac{\rho u^2}{2} . \label{4}
\end{equation}
Then, we take into account that
\begin{equation}
\frac{\partial \mathcal{E}}{\partial \bm{\Pi }}={\bf v}, \qquad
\frac{\partial \mathcal{E}}{\partial {\bf E }}=-{\bf P}, \label{6}
\end{equation}
where {\bf v} is the velocity of the superfluid system.
 Using Eqs. (\ref{4}) and
(\ref{6}), we find the required relation
\begin{equation}
\bm{\Pi }=\rho {\bf v} -\frac{1}{c}{\bf P}\times {\bf B}.
\label{7}
\end{equation}

 In general case the kinematic momentum $\rho{\bf v}$ is of order
 of the generalized momentum ${\bm \Pi}$. But, as it is shown below,
 for the condenser superconductors in a strong magnetic
 field the kinematic momentum is much smaller than the
 generalized one. Consequently, in a strong magnetic field the term
 $\rho{\bf v}$ in Eq. (\ref{7}) can be omitted. Taking into
 account, that in the superfluid system
$\bm{\Pi }=n\hbar \nabla \varphi$, where $n$ is the density of the
particles (we consider the temperature $T=0$), and $\varphi$ is
the phase of the order parameter, we obtain from Eq. (\ref{7}) the
following relation
\begin{equation}
n\hbar \nabla \varphi =-\frac{1}{c}{\bf P}\times {\bf B}.
\label{9}
\end{equation}
Let the system, lying in the $(x,y)$ plane, is subjected with a
uniform magnetic field, directed along the $z$ axis. Taking the
curl of the both parts of Eq. (\ref{9}), we obtain
\begin{equation}
n \hbar\ {\rm curl}_z \nabla \varphi = \frac{H}{c} {\rm div}_2
{\bf P}, \label{10}
\end{equation}
where  ${\rm div}_2 {\bf P} $ is the two-dimensional divergence.
 of the vector {\bf P}.
This quantity taken with its sign changed is equal to the
polarization charge density $\rho_{\rm pol}$. On the other hand,
the left hand side of Eq. (\ref{10}) is nonzero only in case, when
the quantized vortices exist in the system. Then
\begin{equation}
{\rm curl}_z \nabla \varphi = 2\pi \sum_i \delta ({\bf r}-{\bf
r}_i) n_i, \label{11}
\end{equation}
where $n_i=\pm 1$ the upper (low) sign corresponds to the vortices
rotating in the counter-clockwise (clockwise) direction, and the
summation is over the vortex centers.  Integrating  the both parts
of Eq. (\ref{10}) over an arbitrary area,   for vortices of the
same sign one finds
\begin{equation}
\pm  2\pi\hbar n N_v =\frac{H}{c}\int \rho_{\rm pol} d S\equiv
\frac{H}{c} Q_v. \label{12}
\end{equation}
Here $N_v $ is the number of the vortices in the area $S$, and
$Q_v$ is their charge. It follows from this, that vortex charge is
equal to
\begin{equation}
q\equiv \frac{Q_v }{N_v }=\pm 2\pi \frac{\hbar c}{B} n=\pm 2\pi e
\ell_B^2 n. \label{13}
\end{equation}
In the case of the electron-hole pairing in the lowest Landau
level, the density of the pairs $n$ is related with the filling
factors $\nu_e=\nu_h=\nu$ by the formula $\nu=2\pi\ell_B^2 n$.
Thus, in strong magnetic fields in the superfluid phase the
quantized vortices carry an  electrical charge $q=\pm \nu e$.

To clarify how general is the result obtained let us analyze the
behavior of the condenser superconductor in a strong magnetic
field perpendicular to the layers at small filling factors $\nu$,
when the  electron-hole pair gas can be considered as a weakly
interacting Bose gas. Let us consider two 2D conducting layers
separated by the dielectric layer of the width $d$ with the
electron carriers in one layer and the hole carriers in the
adjacent layer.  We consider $m_h\gg m_e$. In a strong magnetic
field  the large difference of the masses $m_e$ and $m_h$ may
result in a situation, when the electron Bohr radius
$a_B^e=\varepsilon \hbar^2/m_e e^2$ is much larger then the
magnetic length $\ell_B=(c\hbar/eB)^{1/2}$, and the hole Bohr
radius $a_B^h=\varepsilon \hbar^2/m_h e^2$ can be smaller or
larger then $\ell_B$. In this case the energy spectrum of the
bounded electron-hole pair, which is formed due to the Coulomb
interaction between spatially separated carriers, was found in
\cite{16}.  The part of the pair energy depending on the momentum
of the pair ${\bm \pi}$ and the velocity ${\bf u}$ is equal to
\begin{equation}\label{2a}
  \Delta{\cal E}=\frac{\pi^2}{2 M_*}+\frac{M_B}{M_*} {\bf u} {\bm \pi}
  -\frac{1}{2} \frac{M_B}{M_*} m_h u^2.
\end{equation}
Here $M_*=M_B+m_h$, the pair effective mass, $M_B$, the "magnetic
mass"
\begin{equation}\label{3a}
  M_B = \frac{4}{\sqrt{2\pi }} \frac{\varepsilon\hbar^2}{e^2
  \ell_B}=\frac{4}{\sqrt{2\pi}} m_h \frac{a_B^h}{\ell_B }.
\end{equation}
Introducing the pair polarizability
\begin{equation}\label{4a}
  \alpha(B)=M_B \frac{c^2}{B^2}=
  \frac{4\varepsilon}{\sqrt{2\pi}} \ell_B^3,
\end{equation}
one can rewrite the correction $\Delta\cal{E}$ in the form
\begin{equation}\label{5a}
  \Delta{\cal E}=\frac{1}{2 M_*}\left({\bm \pi}+ \alpha(B)
  \frac{{\bf E}\times{\bf B}}{c}\right)^2 -\frac{\alpha(B)}{2}
  E^2.
\end{equation}
Analogous expression was obtained in \cite{17} for an electrically
neutral atom in crossed fields for the case of small magnetic
fields. In that case Eq. (\ref{5a}) contains the zero magnetic
field polarizability of the atom $\alpha(0)$ instead of
$\alpha(B)$ and the mass of the atom $M$ instead of the mass of
the pair $M_*$.

Replacing the momentum ${\bm \pi}$ with the operator $-i\hbar
\nabla$,  from Eq. (\ref{5a}) we obtain the Hamiltonian of the
electron-hole pairs. In the low density limit, when the size of
the pair is much smaller then the distance between the pairs and
the exchange effects are inessential, the pairs can be considered
as true bosons. At low temperatures the rarefied Bose gas should
form a superfluid state. The superfluid phase can be described by
the order parameter $\Psi$. The order parameter satisfies the
equation
\begin{eqnarray}\label{6a}
  i\hbar \frac{\partial\Psi}{\partial t}=\frac{1}{2 M_*} \left(
  -i\hbar \nabla +\alpha(B)\frac{{\bf E}\times{\bf B}}{c}\right)^2
  \Psi \cr -\frac{\alpha(B)}{2} E^2 \Psi+\gamma |\Psi|^2 \Psi.
\end{eqnarray}
The last term in the r.h.s. of Eq.(\ref{6a}) describes the
interaction between the pairs. One can show that in the the limit
$d\ll \ell_B$ the interaction constant is equal to
$\gamma=(\pi/2)^{3/2} e^2 d^2/\varepsilon \ell_B$. The vanishing
of the interaction constant at $d=0$ is the consequence of the
exact compensation of the Coulomb forces between the pairs
(compare with \cite{d1}). Presenting the order parameter in the
form $\Psi=|\Psi|e^{i\varphi(r)}$, we obtain from Eq.(\ref{6a})
the velocity of the superfluid component
\begin{equation}\label{7a}
{\bf v}_s=\frac{1}{M_*} \left(\hbar\nabla \varphi +\alpha(B)
\frac{{\bf E}\times{\bf B}}{\hbar c}\right).
\end{equation}

To obtain the dipole momentum of the unit area ${\bf P}$ we take
into account that the r.h.s. of Eq. (\ref{6a}) is the variational
derivative over $\Psi^*({\bf r})$ of the energy functional of the
Gunzburg-Landau type for superconductors. The derivative of that
functional over the electric field ${\bf E}$ taken with the
opposite sing is ${\bf P}$:
\begin{equation}\label{8a}
  {\bf P}=\alpha(B) \left[\left(1-\frac{M_B}{M_*}\right){\bf E}+\frac{\hbar}{M_* c}
  \nabla \varphi\times {\bf B}\right]
  |\Psi|^2.
\end{equation}
The expression (\ref{8a}) can be rewritten in the form
\begin{equation}\label{9a}
{\bf P}=\alpha(B) \Big({\bf E}+\frac{1}{c} {\bf v}_s\times {\bf
B}\Big)
  |\Psi|^2.
\end{equation}

This result means that not only the electric field {\bf E}, but
also the Lorentz force polarizes the pair, acting in opposite
directions on the positive and negative charges of the pair.
However, such a simple expression for ${\bf P}$ is valid within
the linear accuracy in $v_s/c$.

One can note that the expressions (\ref{7a}) and (\ref{8a}) are in
agreement with Eq. (\ref{7}). Actually, if we exclude from Eqs.
(\ref{7a}) and (\ref{8a}) the electric field ${\bf E}$ and take
into account that in the case considered $\rho=m_h |\Psi|^2$, we
arrive to the relation  (\ref{7}).

The polarization charge $\rho_{pol}$  is found by taking the
two-dimensional divergence of the both sides of Eq. (\ref{8a}). To
calculate the derivatives in the r.h.s. of Eq. (\ref{8a}) we take
into account that the relation (\ref{11}) for the ${\rm curl}_z
\nabla \varphi $ takes place. Beside this we omit the terms
containing the quantity $\nabla|\Psi|$. We are not interested here
in the  structure  of the vortex core (we consider it as the
mathematical point), and out of the core the terms containing
$\nabla|\Psi|$ are small at small fields $E$. It allows to replace
$|\Psi|^2$ with the pair density $n$. Finally, we obtain
\begin{eqnarray}\label{10a}
  \rho_{pol}({\bf r})= -\alpha(B) n
  \Bigg[\left(1-\frac{M_B}{M_*}\right) {\rm div}_2 {\bf E}+\cr \sum_i
  2\pi \frac{\hbar B}{M_* c} n_i \delta({\bf r}-{\bf r}_i)\Bigg].
\end{eqnarray}

Thus, the polarization charge in a superfluid system in a magnetic
field can be caused by the polarization of the medium by the
electric field with a nonzero divergence, or by the existence of
the quantized vortices in the system.  It  also follows from Eq.
(\ref{10a}), that the charge of the $i$-th vortex is equal to the
coefficient of $\delta({\bf r}-{\bf r}')$, namely
\begin{equation}\label{11a}
  q= \pm 2\pi\frac{\hbar B}{M_* c} \alpha(B) n.
\end{equation}
Eq. (\ref{11a}) yields the vortex charge for the electron-hole
double layer systems in an arbitrary magnetic fields. The same
result is valid for the electron-electron double layer system with
the substitution $M_*=M_B+2 m_e$.

 In  weak magnetic
 fields ($\ell_B\gg a_B^e$) the polarizability $\alpha=\gamma
 (a_B^e)^3$, where $\gamma\sim 1$, and the effective mass
 $M_*\simeq m_h+m_e\simeq m_h$. Then, the vortex charge is
 equal to
\begin{equation}\label{12a}
  q=\pm \frac{2\pi\gamma}{\varepsilon}
  \left(\frac{a_B^e}{\ell_B}\right)^2 a_B^e a_B^h n e.
\end{equation}
In  high magnetic fields ($a_B^e\gg \ell_B \gtrsim a_B^h$) using
 Eq. (\ref{4a}) for $\alpha(B)$ one obtains
\begin{equation}\label{13a}
  q=\pm 2\pi \ell_B^2 n \frac{M_B}{M_*} e =\pm \frac{M_B}{M_*} \nu
  e.
\end{equation}
Finally, in ultra high fields ($a_B^h\gg \ell_B$) the effective
mass $M_*=M_B(1+\sqrt{2\pi} \ell_B/4 a_B^h)\to M_B$ and the vortex
charge obeys the universal relation $q=\pm \nu e$.

 Let us present here some estimates. In the magnetic field $B= 10$
 T the magnetic length $\ell_B\approx 80$ \AA. In GaAs
 heterostuctures with the dielectric constant $\varepsilon=13$
 the magnetic mass  is very small ($M_B\approx 10^{-28}$ g). Due
 to the smallness of this quantity the vortex charge can be of
 order the electron charge only in superfluid systems with
 the boson  mass of order  the electron mass
 $m_0$. For the electron-hole double layers at $m_h=0.4\
 m_0$ and $m_e=0.067 m_0$ we obtain  $M_B/M_*\approx 0.2$ and the
 vortex charge is equal to $q\approx 0.2\nu e$. For the
 electron-electron double layer system with the same $m_e$ we find $M_B/M_*\approx 0.5$
and the vortex charge $q\approx 0.5 \nu e$
 The derivation of $q$
 from the universal value is connected with
that the ratio $\ell_B/a_B^e\approx 0.8$ and
 the strong inequality
 $\ell_B\ll a_B^E$ does not satisfy  in this case.

The number of the vortices and their spatial distribution depends
on the electric field ${\bf E}$. Integrating the both sides of Eq.
(\ref{7a}) along a certain contour, we find
\begin{equation}\label{14a}
  M_*\oint {\bf v}_s d {\bf l} =  2 \pi \hbar N_v
  -\frac{\alpha(B)}{c} B\oint {\bf E}\times d{\bf l}.
\end{equation}
Here $N_v$ is the number of the vortices inside the contour (we
consider the vortices of the same vorticity). It  follows from
Eq.(\ref{14a}), that the velocity ${\bf v}_s$ can be reduced under
the appearance of the vortices - in such a way we lower the
kinetic energy of the system. When the vortex distribution is
considered as a continuous one, the vortex density $n_v({\bf r})$
can be introduced
\begin{equation}\label{15a}
  N_v=\int n_v({\bf r}) d{\bf r}.
\end{equation}
Putting the r.h.s. of Eq.(\ref{14a}) to zero we find from Eqs.
(\ref{14a}) and (\ref{15a}) the relation
\begin{equation}\label{16a}
  n_v({\bf r})=\frac{\alpha(B) B}{2 \pi \hbar c} {\rm div}_2 {\bf
  E}.
\end{equation}

In the case of a condenser superconductor of a disk shape and for
the electric field directed along the disk radius and independent
of the angle $\varphi$, it is reasonable to assume, that the
vortices situated on the circumferences centered at the disk
center. If the radius of the $i$-th circumference is $\rho_i$ and
the number of the vortices on it is $N_i$, then it follows from
the definition of  $n_v({\bf r})$
\begin{equation}\label{17a}
  N_i =\int_{\rho_i}^{\rho_{i+1}} \int_0^{2 \pi} n_v({\bf r}) r d
  r d \theta.
\end{equation}

This equation determines the relation between the quantities
$\rho_i$ and $N_i$. The values of $\rho_i$ can be found from the
assumption that the average distance between the vortices on a
given circumference (which is equal $2\pi\rho_i/N_i$) coincides
with the average distance $\rho_{i+1}$ - $\rho_i$ between the
vortices belonging to the adjacent circumferences.  It yields
\begin{equation}\label{18a}
  N_i= \frac{2\pi \rho_i}{\rho_{i+1}-\rho_{i}}.
  \end{equation}
Eqs. (\ref{16a})-(\ref{18a}) allow to find the values of $N_i$ and
$\rho_i$ up to a factor of order of unity.

A macroscopic number of the vortices with equal vorticites can
also emerge in the absence of the electric field. It is realized
when besides  the uniform field ${\bf B}_z$ there is an extra
field ${\bf B}_\tau$ with ${\rm div}_2 {\bf B}_\tau\ne 0$ (${\bf
B}_\tau$ is parallel to the plane of the structure). Indeed, one
can show (compare with \cite{18}) that  in the field ${\bf
B}_\tau$ the energy of the pair of  spatially separated electron
and  hole is equal to
\begin{equation}\label{19a}
  {\cal E}=\frac{1}{2M_*} ({\bm \pi}+ \frac{e d}{c} \hat{z}\times
  {\bf B}_\tau)^2.
\end{equation}
One can see that energy (\ref{19a}) differs from the expressions
(\ref{5a}) only by that the  induced dipole momentum $\alpha(B)
{\bf E}$ is replaced with the spontaneous momentum $ed\hat{z}$.
Therefore, the dipole momentum of the unit
area  can be obtained from Eq. (\ref{9a}) 
replacing the induced momentum with the spontaneous one. Then,
taking the divergence of ${\bf P}$, we find
\begin{equation}\label{20a}
  \rho_{pol}({\bf r})=- \frac{\alpha(B) B n}{M_* c} \Bigg[ \frac{e
  d}{c} {\rm div}_2 {\bf B}_\tau+ \sum_i 2 \pi \hbar n _i
  \delta({\bf r}-{\bf r}_i)\Bigg].
  \end{equation}
It follows from this expression that in this case the vortex
charge is equal the value found above and in the continuous limit
the vortex density is $n_v({\bf r})= (ed/2\pi \hbar c) {\rm div}_2
{\bf B}_\tau$.

At nonzero temperatures the charged vortices  will emerge in
condenser superconductors in a fluctuation way,  in similarity
with the same phenomena in  a thin He-II film. The circumstance
that the vortices are charged does not influence in the first
approximation on the thermodynamic features of the system. It is
connected with that the Coulomb correction to interaction between
the vortices falls down much faster (by the power law)
then bare logarithmic interaction between them. The last one, as
is well known, results in a Kosterlitz-Thouless transition. Since
the sign of the vortex charge is in one to one correspondence with
the sign of the vorticity, at temperatures below the
Kosterlits-Thouless temperature the vortex-antivortex pairs should
be electrically neutral. At temperature above the
Kosterlits-Thouless the vortices and antivortices  decouple, and
free electrical charges appear. It reveals itself in a principal
change of the conducting properties of the system under the phase
transition.

This work is supported by the INTAS grant No 01-2344

\end{document}